\documentclass[twocolumn]{revtex4}
\usepackage{amsfonts}
\usepackage{amsmath}
\usepackage{amssymb}
\usepackage{ragged2e}
\usepackage{graphicx}
\usepackage[usenames,dvipsnames]{color}
\definecolor{darkblue}{RGB}{0,0,196}

\begin{document}

\title{Particle species and energy dependencies of freeze-out parameters in high-energy proton-proton collisions.
\vspace{0.5cm}}

\author{Muhammad Waqas$^{1}$\footnote{waqas\_phy313@yahoo.com; waqas\_phy313@ucas.ac.cn},
Guang Xiong Peng$^{1,2}$\footnote{Correspondence: gxpeng@ucas.ac.cn},
Fu-Hu Liu$^{3}$\footnote{fuhuliu@163.com; fuhuliu@sxu.edu.cn},
Muhammad Ajaz$^{4}$\footnote{Correspondence: ajaz@awkum.edu.pk; muhammad.ajaz@cern.ch},
Abd Al Karim Haj Ismail$^{5,6}$\footnote{a.hajismail@ajman.ac.ae}
Khusniddin K. Olimov$^{7}$\footnote{khkolimov@gmail.com},
Abdel Nasser Tawfik$^{8}$\footnote{a.tawfik@fue.edu.eg}
\vspace{0.25cm}}

\affiliation{$^1$School of Nuclear Science and Technology, University of Chinese Academy of Sciences, Beijing 100049, China,
\\
$^2$Theoretical Physics Center for Science Facilities, Institute of High Energy Physics, Chinese Academy of Sciences, Beijing 100049, China
\\
$^3$Institute of Theoretical Physics \& State Key Laboratory of Quantum Optics and Quantum Optics Devices,
Shanxi University, Taiyuan, Shanxi 030006, China
\\
$^4$Department of Physics, Abdul Wali Khan University Mardan, 23200 Mardan, Pakistan
\\
$^5$College of Humanities and Sciences, Ajman University, PO Box 346, United Arab Emirates
\\
$^6$Nonlinear Dynamics Research Center (NDRC), Ajman University, PO Box 346, UAE
\\
$^7$ Physical-Technical institute of Uzbekistan Academy of Sciences, Tashkent 100084, Uzbekistan
\\
$^8$Future University in Egypt (FUE), 5th Settlement, 11835 New Cairo, Egypt}
\begin{abstract}

\vspace{0.5cm}

\noindent {\bf Abstract:} We used blast wave model with Tsallis statistics to analyze the experimental data measured by ALICE Collaboration in proton-proton collisions at Large Hadron Collider and extracted the related parameters (kinetic freeze-out temperature, transverse flow velocity  and kinetic freeze-out volume of emission source) from transverse momentum spectra of the particles. We found that the kinetic freeze-out temperature and kinetic freeze-out volume are mass dependent. The former increase while the latter decrease with the particle mass which is the evidence of a mass as well as volume differential kinetic freeze-out scenario. Furthermore we extracted the mean transverse momentum and initial temperature by an indirect method and observed that they increase with mass of the particles. All the above discussed parameters are observed to increase with energy. Triton ($t$), hyper-triton (${^3_{\bar\Lambda} H}$) and helion (${^3 He}$) and their anti-matter are observed to freeze-out at the same time due to isospin symmetry.
\\

\end{abstract}

\maketitle

\section{Introduction}

Although the ultra-relativistic heavy ion central collisions were
envisioned to produce a de-confined state of partons known as Quark-Gluon Plasma (QGP) \cite{1, 2}, the remarkable collision energies at Large
Hadron Collider (LHC) has brought up emerging tasks in specifying the
proton-proton ($pp$) collisions for understanding a possible formation of QGP droplets in these hadron collisions. Many signatures of QGP including
strangness enhancement \cite{3, 4}, hardening of transverse momentum ($p_T$) spectra \cite{5, 6} and thermal effective temperature being comparable to that observed in heavy ion collisions and degree of collectivity \cite{7} etc have been observed in $pp$ collisions at LHC. In addition, the high multiplicity in $pp$ collisions at LHC also provides the opportunity to study the de-confined matter under the extreme conditions of high temperature and/or energy density. The initial energy density results in high pressure gradient, which leads the fireball to expand. The particles produced experience both elastic as well as inelastic interaction with each other. Recently, in $pp$ collisions, the suppression of $K^{*0}/K$ ratio as a function of charged particle multiplicity was observed which implies the presence of hadronic phase in high multiplicity $pp$ collisions with non-zero life-time \cite{5}. This hadronic phase is said to be the phase between the chemical and thermal/kinetic freeze-out. The freeze-out hyper-surface, where the inelastic collisions stop and the particle ratios get fixed is named as chemical freeze-out, after which the particles experience elastic collisions with each other till the kinetic freeze-out, where the mean free path of the particles become larger than the size of the system. The  ($p_T$) spectra of the particles is used to determine the kinetic freeze-out hyper-surface.
Freeze-out is very complicated process and have contradiction in different literature. According to \cite{8}, there is single freeze-out scenario for all the particles, however according to \cite{9, 10} a double freeze-out scenario exists where non-strange and strange particles decouple separately from the system. Some literature \cite{11, 12, 13} claim the multiple kinetic freeze-out scenario. One of our recent works \cite{14} also observed a triple kinetic freeze-out scenario. In fact, it is hard to say that which scenario is correct. Furthermore, the freeze-out parameters such as kinetic freeze-out temperature ($T_0$), transverse flow velocity ($\beta_T$) and kinetic freeze-out volume ($V$) also depends on other things such as collision energy \cite{13, 15, 16}, centrality \cite{14, 16, 17, 18, 19, 20, 21, 22, 23, 24}, production cross-section  \cite{14, 17}, multiplicity \cite{25} and rapidity \cite{26}. We wonder whether there is also a role of quarks or nucleons
coalescence, and isospin symmetry in the freeze-out of particles or not.

In the present work, we will study behavior of the freeze-out parameters in $pp$ collisions at LHC, which are very important. We believe that the specific variation in excitation functions of the freeze-out parameters, especially kinetic freeze-out temperature and kinetic freeze-out volume, are associated to the critical end point (CEP) of the phase transition from hadronic matter to QGP happened in central  nucleus-nucleus collisions, where the specific change renders the appearances of saturation, minimum, maximum, asymptotic line etc.
In case of high multiplicity $pp$ collisions, even if the collision is minimum-bias, the specific variation of some parameters are assumed
to be comparable with those in AA collisions, where it is expected that the quarks degree of freedom in minimum-bias $pp$ collisions play initially a main role at the energy of specific change. Because of small system and small number of secondaries in minimum-bias $pp$ collisions, there is no expectation of QGP to be formed. The minimum parameters are related to the soft point of equation of state (EOS) of hadronic matter or QGP phase transition.

In this work we will analyze identified and strange particles, and light nuclei produced in $pp$ collisions at different center-of-mass energies at LHC, and will extract $T_0$, $\beta_T$ and $V$ from their $p_T$ spectra. In addition, we will also extract the $<p_T>$ and $T_i$ by an indirect way.

\section{Formalism and method}

The Hagedorn function \cite{27, 28, 28a, 28b} or quantum chromodynamics (QCD) calculus \cite{29, 30, 31} describes that the high $p_T$ region is generally contributed by the hard scattering process. The Hagedorn function is an inverse power law, and in our opinion it has at least three revisions \cite{32, 33, 34, 35, 36, 37, 38, 38a}.
We shall skip to discuss the Hagedorn function and its revisions, not containing the freeze-out parameters such as $T_0$, $\beta_T$ and $V$ directly. Although the main contribution fraction to the parameters is from the low $p_T$ region, that from high $p_T$ region is also interesting for us.

In order to extract the freeze-out parameters such as $T_0$, $\beta_T$ and $V$, the $p_T$ spectra should be analyzed. Although various
functions can be chosen to describe the mentioned function, the blast wave model with Tsallis statistics
which takes into account the collective flow in both the longitudinal and transverse directions in high energy collisions is more convenient because blast wave model is easy to use for the extraction of
$T_0$, $\beta_T$ and $V$, and on the other hand, the Tsallis distribution can cover the spectra as wide as possible. Many authors have adopted this formula to fit the data in relativistic high energy
collisions \cite{39, 40, 40a 41, 41a, 42}. It is relatively trivial if we change the particle emission source from Boltzmann
to Tsallis distribution.

According to \cite{8, 43, 44}, the blast wave model with Tsallis statistics results in the $p_T$ distribution to be
\begin{align}
f(p_T)=&\frac{1}{N}\frac{\mathrm{d}N}{\mathrm{d}p_\mathrm{T}}=C \frac{gV}{(2\pi)^2} p_T m_T \int_{-\pi}^\pi d\phi\int_0^R rdr \nonumber\\
& \times\bigg\{{1+\frac{q-1}{T_0}} \bigg[m_T  \cosh(\rho)-p_T \sinh(\rho) \nonumber\\
& \times\cos(\phi)\bigg]\bigg\}^\frac{-1}{q-1}
\end{align}
where $N$ is the number of particles, $\rho=\tanh^{-1} [\beta(s)(r/R)^n]$ represents the flow profile
which grows as \textit{(n)}-th power from zero at the center of collisions to $\beta_S$ at the hard spherical edge \textit{(R)}
along the transverse direction. $\beta_T=(2/R^2)\int_0^R r\beta(r)dr=2\beta_S/(n_0+2)=2\beta_S/3$, where
$n_0$ is taken to be 1 \cite{45}. $g$ is the degeneracy factor which changes from particle to particle based on
$g_n$=2$S_n$+1 ($S_n$ is the spin of the particle), \textit{V} is the freeze-out volume, $m_T=\sqrt{p_T^2+m_0^2}$ is
the rest mass, $\phi$ is the emission angle in thermal source rest frame, and $q$ is the entropy index which
characterizes the degree of non-equilibrium of the produced system.

In fact, the structure of $p_T$ spectra is very complicated. There are several $p_T$ regions and each $p_T$
region corresponds to different mechanisms which are discussed in detail in previous work \cite{46}. In the present
work, the blast wave model with Tsallis distribution is used to describe the soft excitation process. However
the fit in high $p_T$ region is not good in some cases; then we have to use the two-component fit in which
the second component describes the hard scattering process. In case of considering simultaneously the hard scattering, the main contributor of particle production is still the soft processes. The second component has the same form as the first
component, and the two components are then structured as

\begin{align}
f_0(p_T)=kf(p_T,T_{01},\beta_{T1})+(1-k)f(p_T, T_{02}, \beta_{T2}),
\end{align}
\textit{k} in Eq. (2) is the contribution fraction of the two-component. We can also use the usual step function in order to structure
the two-component TBW model, according to Hagedorn model function \cite{27}
\begin{align}
f(p_T)=A_1\theta(p_1-p_T) f(p_T,T_{01},\beta_{T1}) \nonumber\\+ A_2 \theta(p_T-p_1)f(p_T,T_{02},\beta_{T2}),
\end{align}
where $A_1$ and $A_2$ are constants.

In order to extract $T_0$, $\beta_T$ in two component TBW model, Eq. (2) and (3) can be used
\begin{align}
T_0=kT_{01}+(1-k)T_{02},
\end{align}

and
\begin{align}
\beta_T=k\beta_{T1}+(1-k)\beta_{T2},
\end{align}
In addition, we considered the normalization constant, and the real fitted kinetic freeze-out
volume should be $V_1$=$N_1$/$\bar V_1$/$k$ and $V_2$=$N_2$/$\bar V_2$/$(1-k)$, then as we know
that the volume component has the additive property, and it can be structured as $V$=$V_1$+$V_2$.

\section{Results and discussion}

\subsection{Comparison with data}

Figure 1 shows the $p_{T}$ spectra  ($1/N_{ev}$ $d^2N$/$dp_T$\textit{dy}, $1/N_{NSD}$ $d^2N$/$dp_T$\textit{dy},
$d^2N$/$dp_T$dy, $1/N_{INEL}$ $d^2N$/$dp_T$dy or $1/N_{ev}$ (1/2$\pi$ $p_T$)$d^2N$/$dp_T\textit{dy}$
of different particles in in-elastic (INEL) or non-single diffractive (NSD) $pp$ collisions at different energies. Panel (a) shows the $p_T$ spectra of the particles in INEL $pp$ collisions
at $\sqrt{s}$=0.9 TeV. These particles include $\pi^+$, $\pi^-$, $K^+$ , $K^-$, $p$, $\bar p$, $K^0_S$, $\Lambda+\bar \Lambda$,
$\Xi^-$, $d$, $\bar d$, $^3_\Lambda H$, $^3_\Lambda \bar H$, $^3{He}$ and $\bar {^3He}$. The
symbols represent the measured experimental data of ALICE Collaboration at LHC,
while the curves are the result of our fitting by Eq. (1). Different symbols are used to display different particles and they are
labeled inside the panel. The data of $\pi^+$, $\pi^-$, $K^+$, $K^-$, $p$ and $\bar p$
are taken from ref. \cite{47} in the mid-rapidity range $|y|<0.5$, while the data for $K^0_S$ and
$\Xi^-$ at $0<|y|<2$ are taken from ref. \cite{48}, and the data for
$\Lambda+\bar \Lambda$, $d$, $\bar d$, $^3_\Lambda H$, $^3_\Lambda \bar H$, $^3{He}$ and
$\bar {^3He}$ in $|y|<0.5$ are taken from ref. \cite{49}.

Panel (b) displays the $p_T$ spectra of $\pi^++\pi^-$, $K^++K^-$, $p+\bar p$, $\phi$, $\Lambda+\bar \Lambda$,
$\Xi^-$, $\bar \Xi^+$, $\Omega^-+\bar \Omega^+$, $d$, $\bar d$, $^3_\Lambda H$, $^3_\Lambda \bar H$, $^3{He}$ and $\bar {^3He}$ at 2.76 TeV in pp collisions. The symbols represent the measured experimental data of ALICE Collaboration at LHC,
while the curves are the result of our fitting by using Eq. (1). Different particles are represented by different symbols, and they are
labeled inside the panel. The data of $\pi^++\pi^-$, $K^++K^-$, $p+\bar p$ in mid-pseudo-rapidity
$|\eta|<0.2$ are taken from the ref \cite{50}, while the data for $\phi$ at $2.96<|y|<3.53$ are taken from ref. \cite{51}
and for $\Xi^-$, $\bar \Xi^+$ and $\Omega^-+\bar \Omega^+$ $0<|y|<0.5$, the data
is taken from ref \cite{52}. The data for $\Lambda+\bar \Lambda$, $d$, $\bar d$, $^3_\Lambda H$, $^3_\Lambda \bar H$, $^3{He}$ and
$\bar {^3He}$ in $|y|<0.5$ are taken from ref.\cite{51}.

In panel (c), the $p_T$ spectra of $\pi^++\pi^-$, $K^++K^-$, $p+\bar p$, $K^0_S$, $(K^*+\bar K^*)/2$, $\phi$, $\Lambda$,
$\Xi^-$, $d$, $\bar d$, $t$, $\bar t$, $^3_\Lambda H$, $^3_\Lambda \bar H$, $^3{He}$ and $\bar {^3He}$ produced
in $pp$ collisions at 7 TeV are presented. The symbols represent the measured experimental data of ALICE Collaboration at LHC,
while the curves are the result of our fitting by using Eq. (1). Different symbols are used to display different particles and they are
labeled inside the panel. The data of $\pi^++\pi^-$, $K^++K^-$, $p+\bar p$ $|y|<0.5$ are taken from ref \cite{53}, while the data for $K^0_S$ in $0<|y|<2$ are taken from ref. \cite{50}.
$(K^*+\bar K^*)/2$ and $\phi$ data are taken from ref. \cite{54}, and $\Lambda$ and for $\Xi^-$ is from ref. \cite{50}.
The data for $d$, $\bar d$, $t$, $\bar t$, $t$, $\bar t$, $^3{He}$ and $\bar {^3He}$ are taken from ref. \cite{55} and for $^3{He}$ and $\bar {^3He}$ are taken from ref. \cite{51}.

\begin{figure*}[htb!]
\begin{center}
\hskip-0.153cm
\includegraphics[width=15cm]{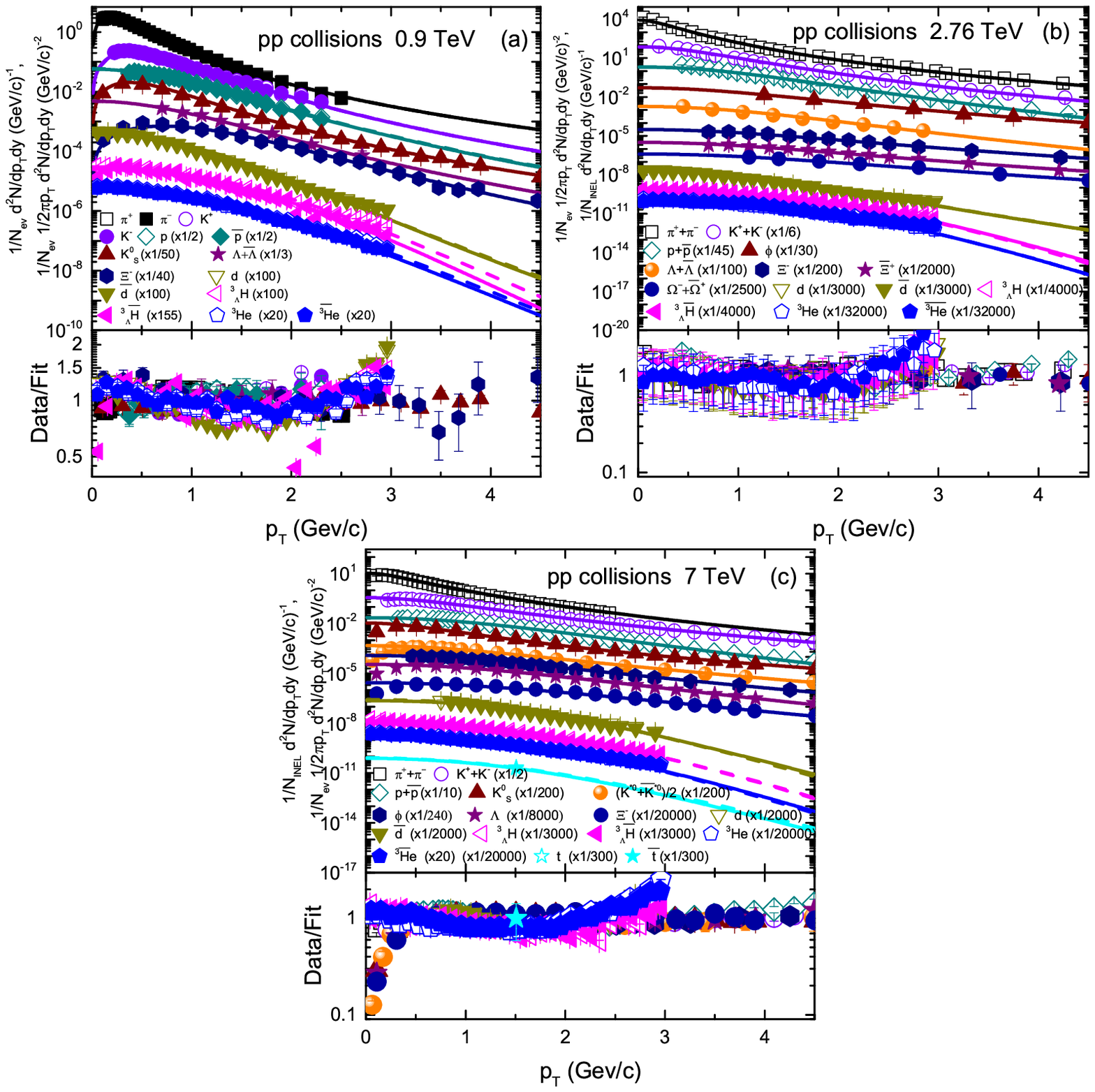}
\end{center}
Fig. 1. Transverse momentum spectra of particles. Panel (a) shows the spectra of $\pi^+$, $\pi^-$, $K^+$ , $K^-$, $p$, $\bar p$, $K^0_S$, $\Lambda+\bar \Lambda$,
$\Xi^-$, $d$, $\bar d$, $^3_\Lambda H$, $^3_\Lambda \bar H$, $^3{He}$ and $\bar {^3He}$ produced in $pp$ collisions at $\sqrt{s}$=0.9 TeV. The symbols represent the experimental data measured by the ALICE Collaboration. Panel (b) represents the $p_T$ spectra of $\pi^++\pi^-$, $K^++K^-$, $p+\bar p$, $\phi$, $\Lambda+\bar \Lambda$, $\Xi^-$, $\bar \Xi^+$, $\Omega^-+\bar \Omega^+$, $d$, $\bar d$, $^3_\Lambda H$, $^3_\Lambda \bar H$, $^3{He}$ and $\bar {^3He}$ at $\sqrt{s}$=2.76 TeV, while panel (c) represents the $p_T$ spectra of $\pi^++\pi^-$, $K^++K^-$, $p+\bar p$, $K^0_S$, $(K^*+\bar K^*)/2$, $\phi$, $\Lambda$,
$\Xi^-$, $d$, $\bar d$, $t$, $\bar t$, $^3_\Lambda H$, $^3_\Lambda \bar H$, $^3{He}$ and $\bar {^3He}$ in $pp$ collisions at $\sqrt{s}$=7 TeV. The curves are our fitted results by Eq. (1). The data used in panel (a) is taken from ref. \cite{47,48,49}, panel (b) from ref. \cite{50, 51, 52} and panel (c) from ref. \cite{50, 51, 53, 54, 55}. Each panel is followed by its corresponding ratios of Data/Fit to show the deviation of curve from the data.
\end{figure*}

In order to show the clear representation of the spectras, some spectra are scaled by some factor, such as
in panel (a) $p$ and $\bar p$ are multiplied by the factor 1/2, $K^0_S$, $\Lambda+\bar \Lambda$
and $\Xi^-$ are multiplied by the the factor 1/50, 1/3 and 1/40 respectively,
while the spectra of $d$, $\bar d$ and $^3_\Lambda H$ are multiplied by 100, the
spectra of $^3_\Lambda \bar H$ are multiplied by 155, and $^3{He}$ and $\bar {^3He}$
are multiplied by 20. In panel (b), the spectra of $K^++K^-$ and $p+\bar p$ are scaled by 1/6 and 1/45 respectively,
while the spectra of $\phi$, $\Lambda+\bar \Lambda$, $\Xi^-$, $\bar \Xi^+$, and $\Omega^-+\bar \Omega^+$,
are scaled by 1/30, 1/100, 1/200, 1/2000 and  1/2500 respectively. The spectra of $d$ and $\bar d$,
$^3_\Lambda H$ and $^3_\Lambda \bar H$, and $^3{He}$ and $\bar {^3He}$ are scaled by 1/3000, 1/4000 and 1/32000
respectively. In panel (c), the spectra of $K^++K^-$ and $p+\bar p$ are scaled by 1/2 and 1/10 respectively,
the spectra of $K^0_S$ and $(K^*+\bar K^*)/2$ are scaled by 1/200, $\phi$, $\Lambda$ and
$\Xi^-$ is scaled by 1/240, 1/8000 and 1/20000 respectively, while the spectra of $d$ and $\bar d$, $t$ and $\bar t$, $^3_\Lambda H$ and $^3_\Lambda \bar H$, and $^3{He}$ and $\bar {^3He}$ are scaled by 1/2000, 1/300, 1/3000 and 1/20000 respectively.

The values of fitting parameters ($T_0$, $\beta_T$ and $V$) with the normalization constant $N_0$, $\chi^2$
and number of degree of freedom (ndof) are listed in table 1. In order to show the deviation between the
experimental data and our fit function curve, each panel is followed by the ratio of its data/fit. The values
of $\chi^2$ in table 1 correspond to the quality of fit. The better fitting results in smaller $\chi^2$, and the
results will be closer to experimental results. It can be seen that the blast wave model with Tsallis
statistics fits satisfactorily the experimental data measured by the ALICE Collaboration at LHC. In most
cases, the good approximate descriptions of the model results for the experimental data can be seen,
however in a few cases, the fit is not so good which results in large dispersion between the data
and the curve, and then naturally $\chi^2$ is larger. {\bf We would like to point out that
the larger $\chi^2$ in a few cases is due to two reasons. (1) the statistics is low
at some of the last points in some cases. (2) We have used a single component function in the present work.
If we would use a two component function, the situation would be different, but the second component from the high
$p_T$ region contributes slightly to the parameters, In deed, we do not needed to consider the second
component in the present work.}
\subsection{Tendencies of parameters}

In order to study the tendency of parameters, the dependences of
$T_0$, $\beta_T$ and $V$ on $m_0$ and collision energy are shown in
Fig. 2. Panel (a) shows the dependence of $T_0$ on $m_0$ and collision energy. The symbols from left to right displays the dependence of $T_0$ on the $m_0$, while from up to downwards is its dependence on
energy. Different symbols in the legends inside the panel are used for the representation of different energies. The symbols with upper filled portion represents the particles while down filled symbols represents the anti-particles. One can see that the kinetic freeze-out
temperature increases with particle mass. The more massive the particle is, the higher is the kinetic freeze-out temperature, which supports the differential freeze-out scenario, revealing that massive particles freeze-out earlier from the system. Furthermore, we observed that there is an effect of isospin symmetry at high energies on $T_0$, which occurs in particles of nearly identical masses where an up quark is replaced by a down quark, and therefore $t$, $^3_\Lambda H$ and $^3{He}$ are observed to freeze-out at the same time. This result obtain suggests
that heavier particles freeze-out earlier as compared
to the lighter ones and the $q$ values obtained listed in the
table 1 suggests that, on the whole, the heavier particles
seem to be closer to equilibrium (lower $q$ values) as compared
to the lighter pions and kaons (higher q values). {\bf In this work,
$q$ is very small in all cases. In our opinion, $q$$<$1.15 renders an approximate equilibrium, which means that blast wave model with Boltzmann Gibbs statistics
is also applicable.}

$T_0$ is observed to increase with increasing the collision energy. This is due to
the reason that the collision is extremely harsh at high energies and large amount
of energy is deposited per particle in the system, which results in larger $T_0$.

\begin{figure*}[htb!]
\begin{center}
\hskip-0.153cm
\includegraphics[width=15cm]{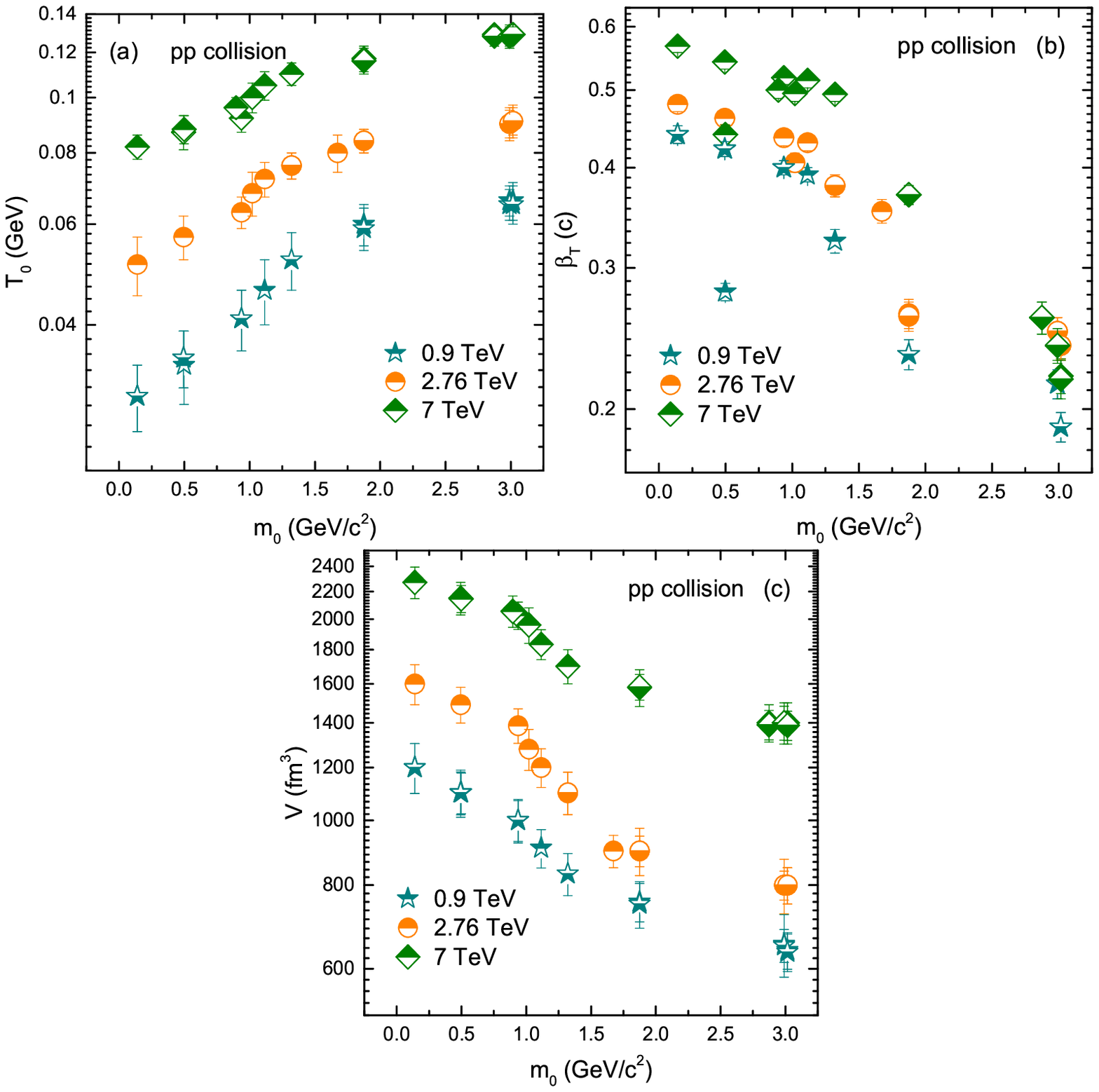}
\end{center}
Fig. 2. Dependence of (a) $T_0$, (b) $\beta_T$ and (c) $V$ on $m_0$ and collision energy.
\end{figure*}

It is noteworthy that during the fitting procedures
the physical restrictions on the parameter values have
been imposed: $T_0$ was allowed to vary between 50 to
300 MeV, and $q$ was restricted to the range from 1.001
to 1.15. Actually, we have given the plausible physical
interpretation of the results obtained to the best of our
knowledge.

Panel (b) is similar to panel (a), but it shows the dependence of $\beta_T$ on $m_0$ and collision energy. The symbols from left to right displays the dependence of $\beta_T$ on the $m_0$, while from up to downwards is its dependence on energy. Different symbols are used for the representation of different energies inside the panel in the legends. The symbols with upper filled portion represents the particles, while down filled symbols represents the anti-particles. We observed that the transverse flow velocity decreases as the particle mass increases. In addition $K^0_S$ is observed to have small $\beta_T$ than some of the heavy particles, which is not understood in the present work. Furthermore, at present, we find that there is no effect of isospin symmetry of the flow velocity.

We also observed that $\beta_T$ increases with the increase of collision energy. This is due to the reason that at higher energies large amount of energy is transferred in system due to high squeeze, which results in more rapid expansion of the fireball and corresponds to larger flow velocity. There is a clear separation among $\beta_T$ values of particles produced in proton-proton collisions at $\sqrt{s}$=0.9, 2.76, and 7 TeV. The largest and lowest $\beta_T$ values are observed in proton-proton collisions at the largest $\sqrt{s}$=7 TeV and lowest $\sqrt{s}$=0.9 TeV collision energies, respectively, while $\beta_T$ values for intermediate energy $\sqrt{s}$=2.76 TeV come in between. This can be explained by that the larger collision energies result in higher pressure gradients
in collision zone and, hence, larger velocities of produced particles flying out from this zone at expansion stage.

Panel (c) is also similar to panel (a), but $V$ dependence on $m_0$
and collision energy is presented in it. The symbols from left to right displays the dependence of $V$ on the $m_0$, while from up to downwards is its dependence on energy. Different symbols are used for the representation of different energies. The symbols with upper filled portion represents the particles, while down filled symbols represents the anti-particles.
One can see that $V$ decreases with particle mass.
The more massive the particle is, the smaller is the parameter $V$, which supports the volume differential freeze-out scenario and indicates that massive particles freeze-out earlier from the system. It may also reveal that there is a separate freeze-out surface for each particle. Furthermore, $V$ for $t$, $^3_\Lambda H$ and $^3{He}$, and $\bar t$, $^3_\Lambda \bar H$, and $\bar {^3He}$ is the same due to isospin symmetry at high energies, which means that they freeze-out at
the same time.

\begin{figure*}[htb!]
\begin{center}
\hskip-0.153cm
\includegraphics[width=15cm]{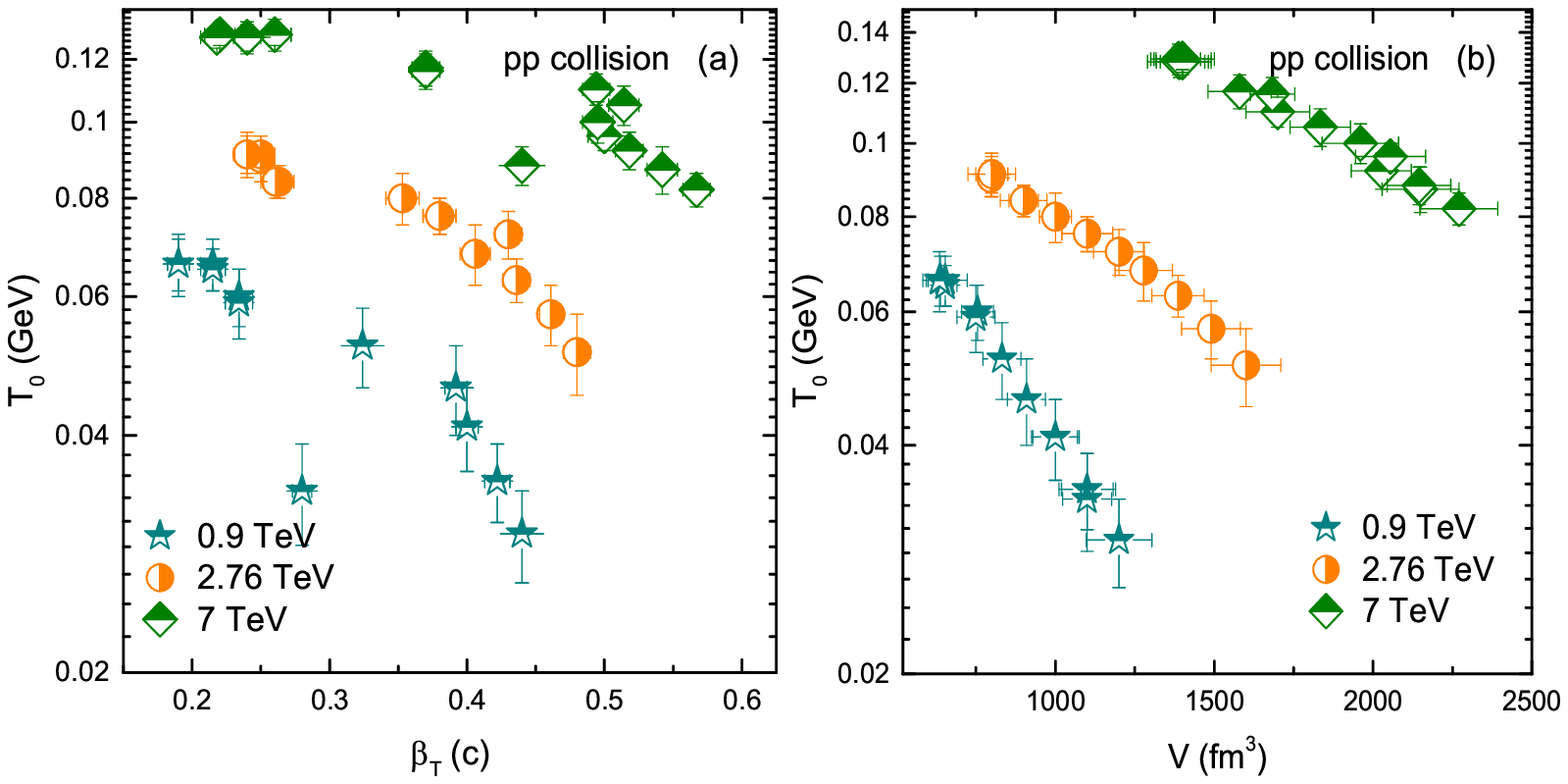}
\end{center}
Fig. 3. Correlation of (a) $T_0$ and $\beta_T$, and (b) $T_0$ and $V$ .
\end{figure*}

\begin{figure*}[htb!]
\begin{center}
\hskip-0.153cm
\includegraphics[width=15cm]{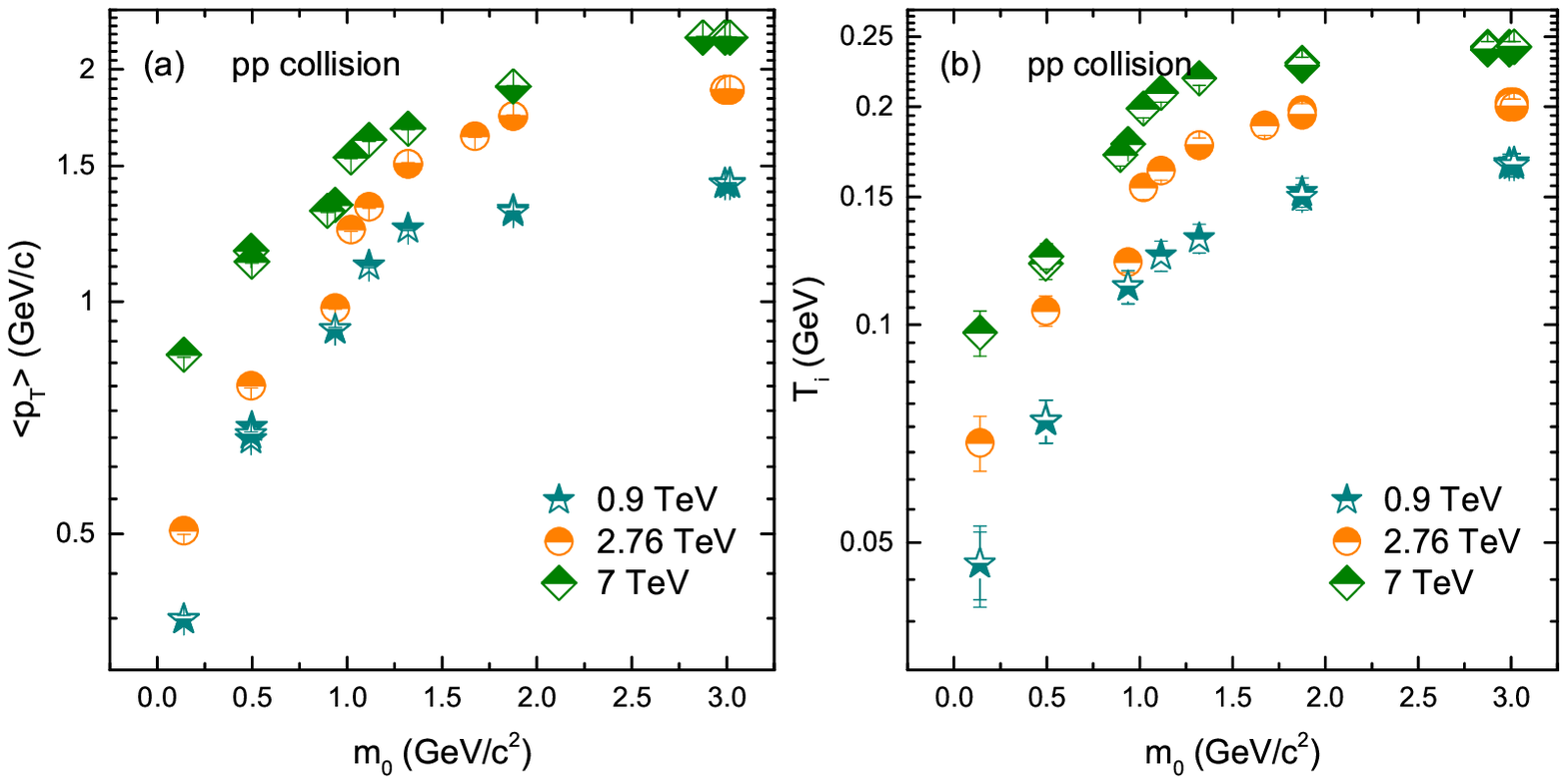}
\end{center}
Fig. 4. Dependence of (a) $<p_T>$ and (b) $T_i$ on $m_0$ and collision energy.
\end{figure*}

Our results show that $V$ increases with increasing the
collision energy. There are two reasons for this. (1) As the collision energy increases, it leads to long evolution time and then larger partonic system, which results in larger $V$. (2) Higher energies corresponds to large multiplicity, which naturally results in larger $V$ if the density saturates.
\begin{table*}
{\scriptsize Table 1. Values of free parameters $T_0$ and
$\beta_T$, V and q, normalization constant ($N_0$),
$\chi^2$, and degrees of freedom (dof) corresponding to the curves
in Fig. 1. \vspace{-.50cm}
\begin{center}
\begin{tabular}{cccccccccccc}\\ \hline\hline
Collisions                  & Particle         & $T_0$           & $\beta_T$        & $V (fm^3)$  &    $q$          & $N_0$                    & $\chi^2$/ dof \\ \hline
Fig. 1                      & $\pi^+$          &$0.030\pm0.004$  & $0.440\pm0.011$  & $1200\pm103$ & $1.14\pm0.005$ & $5\pm0.4$                & 0.3/28\\
   $pp$                       & $K^+$            &$0.035\pm0.004$  & $0.422\pm0.009$  & $1100\pm89$  & $1.112\pm0.005$& $0.62\pm0.04$            & 0.006/22\\
   0.9 TeV                  & $p$              &$0.041\pm0.005$  & $0.400\pm0.008$  & $1000\pm75$   & $1.112\pm0.005$& $0.14\pm0.05$            & 0.005/19\\
                            & $\pi^-$          &$0.030\pm0.004$  & $0.440\pm0.011$  & $1200\pm103$ & $1.14\pm0.005$ & $5\pm0.3$                & 0.5/28\\
                            & $K^-$            &$0.035\pm0.004$  & $0.422\pm0.009$  & $1100\pm81$  & $1.112\pm0.005$& $0.62\pm0.04$            & 0.007/22\\
                            & $\bar p$         &$0.041\pm0.005$  & $0.400\pm0.008$  & $1000\pm70$   & $1.112\pm0.005$& $0.12\pm0.06$            & 0.05/19\\
                            & $K^0_S$          &$0.034\pm0.005$  & $0.280\pm0.007$  & $1100\pm77$  & $1.125\pm0.004$& $3\pm0.5$                & 0.002/16\\
               & $\Lambda+\bar \Lambda$        &$0.046\pm0.006$  & $0.392\pm0.008$  & $909\pm60$   & $1.115\pm0.007$& $0.019\pm0.003$          & 0.6/3\\
                            & $\Xi^-$          &$0.052\pm0.006$  & $0.324\pm0.011$  & $832\pm60$ & $1.08\pm0.006$  & $0.09\pm0.005$            & 1/16\\
                            & $d$              &$0.060\pm0.005$  & $0.234\pm0.010$  & $755\pm50$ & $1.045\pm0.004$ & $3.3\times10^{-6}\pm4\times10^{-7}$       & 5.8/25\\
                            & $\bar d$         &$0.059\pm0.005$  & $0.234\pm0.010$  & $750\pm56$ & $1.045\pm0.006$ & $3.6\times10^{-6}\pm4\times10^{-7}$       & 8.4/25\\
               & ${^3_\Lambda H}$              &$0.066\pm0.005$  & $0.215\pm0.009$  & $653\pm70$ & $1.025\pm0.005$ & $7.2\times10^{-7}\pm5\times10^{-8}$       & 4/25\\
               & ${^3_\Lambda \bar H}$         &$0.065\pm0.004$  & $0.215\pm0.009$  & $650\pm40$ & $1.023\pm0.005$ & $4.2\times10^{-7}\pm4\times10^{-8}$      & 10.9/25\\
               & ${^3 He}$                     &$0.066\pm0.005$  & $0.190\pm0.007$  & $639\pm40$ & $1.03\pm0.006$  & $3\times10^{-7}\pm5\times10^{-8}$         & 18/25\\
               & $\bar {^3He}$                 &$0.066\pm0.005$  & $0.190\pm0.008$  & $635\pm41$ & $1.03\pm0.005$  & $3\times10^{-7}\pm5\times10^{-8}$         & 10/25\\
\hline
Fig. 1                      & $\pi^++\pi^-$    &$0.051\pm0.006$  & $0.480\pm0.010$  & $1600\pm110$ & $1.115\pm0.003$ &$1544\pm123$             & 12/27\\
   $pp$                       & $K^++K^-$      &$0.057\pm0.005$  & $0.461\pm0.008$  & $1490\pm92$  & $1.1\pm0.05$    &$308\pm36$               & 3/21\\
   2.76 TeV                 & $p+\bar p$       &$0.063\pm0.004$  & $0.436\pm0.010$  & $1386\pm82$  & $1.067\pm0.008$ & $69\pm5$                & 15/23\\
                            & $\phi$           &$0.068\pm0.006$  & $0.406\pm0.011$  & $1278\pm90$  & $1.115\pm0.006$& $3.6\pm0.05$             & 2/2\\
               & $\Lambda+\bar \Lambda$        &$0.072\pm0.005$  & $0.430\pm0.010$  & $1200\pm80$  & $1.115\pm0.007$& $0.22\pm0.06$            & 0.001/4\\
                            & $\Xi^-$          &$0.076\pm0.004$  & $0.380\pm0.012$  & $1100\pm80$  & $1.12\pm0.06$  & $0.056\pm0.006$          & 8/8\\
                            & $\bar \Xi^+$     &$0.076\pm0.004$  & $0.380\pm0.012$  & $1100\pm60$  & $1.12\pm0.06$  & $0.0114\pm0.004$         & 8/8\\
           & $\Omega^-+\bar \Omega^+$          &$0.080\pm0.006$  & $0.353\pm0.012$  & $1000\pm70$  & $1.11\pm0.004$ & $1.4\times10^{-3}\pm4\times10^{-4}$     & 11/2\\
                            & $d$              &$0.084\pm0.004$  & $0.263\pm0.011$  & $900\pm50$   & $1.04\pm0.005$ & $4.8\times10^{-5}\pm5\times10^{-6}$     & 24/25\\
                            & $\bar d$         &$0.084\pm0.004$  & $0.261\pm0.011$  & $900\pm47$   & $1.04\pm0.005$ & $4.6\times10^{-5}\pm5\times10^{-6}$     & 29/25\\
               & ${^3_{\bar\Lambda} H}$        &$0.090\pm0.006$  & $0.250\pm0.010$  & $800\pm73$   & $1.004\pm0.0005$& $5.4\times10^{-6}\pm4\times10^{-7}$    & 28/25\\
               & ${^3_{\bar\Lambda} \bar H}$   &$0.090\pm0.005$  & $0.250\pm0.010$  & $797\pm75$   & $1.005\pm0.0004$& $5.2\times10^{-6}\pm4\times10^{-7}$    & 25/25\\
               & ${^3 He}$                     &$0.091\pm0.006$  & $0.240\pm0.009$  & $800\pm40$   & $1.0012\pm0.006$& $3\times10^{-6}\pm4\times10^{-7}$      & 26/25\\
               & $\bar{^3 He}$                 &$0.091\pm0.005$  & $0.240\pm0.010$  & $800\pm50$   & $1.0012\pm0.005$& $3\times10^{-6}\pm4\times10^{-7}$      & 26/25\\
\hline
Fig. 1                      & $\pi^++\pi^-$    &$0.082\pm0.004$  & $0.567\pm0.010$  & $2270\pm123$ & $1.1\pm0.03$    &$3.8\pm0.4$              & 5/33\\
   $pp$                       & $K^++K^-$        &$0.087\pm0.006$  & $0.542\pm0.011$  & $2150\pm121$ & $1.115\pm0.005$ &$0.8\pm0.05$             & 0.04/37\\
   7 TeV                    & $p+\bar p$       &$0.092\pm0.005$  & $0.518\pm0.010$  & $2026\pm95$  & $1.06\pm0.007$  &$0.21\pm0.05$            & 0.006/35\\
                            & $K^0_S$          &$0.088\pm0.005$  & $0.440\pm0.012$  & $2145\pm100$ & $1.15\pm0.06$   &$1.9\pm0.3$              & 4/16\\
               & ($K^{*0}+\bar K^{*0})/2$      &$0.096\pm0.004$  & $0.500\pm0.012$  & $2055\pm110$ & $1.11\pm0.006$  &$0.065\pm0.005$         & 0.01/16\\
                            & $\phi$           &$0.100\pm0.006$  & $0.495\pm0.012$  & $1960\pm120$ & $1.09\pm0.004$  &$0.07\pm0.004$           & 10/19\\
                            & $\Lambda$        &$0.105\pm0.006$  & $0.514\pm0.011$  & $1834\pm95$  & $1.065\pm0.006$ &$0.8\pm0.03$             & 0.2/16\\
                            & $\Xi^-$          &$0.110\pm0.005$  & $0.494\pm0.010$  & $1700\pm100$ & $1.073\pm0.004$ & $0.1\pm0.04$            & 19/16\\
                            & $d$              &$0.117\pm0.006$  & $0.370\pm0.010$  & $1580\pm100$ & $1.001\pm0.0005$& $4\times10^{-4}\pm6\times10^{-5}$     & 56/16\\
                            & $\bar d$         &$0.116\pm0.006$  & $0.370\pm0.010$  & $1583\pm70$  & $1.006\pm0.0005$& $3.7\times10^{-4}\pm5\times10^{-5}$   & 37/15\\
                            & $t$              &$0.129\pm0.004$  & $0.260\pm0.011$  & $1400\pm90$  & $1.01\pm0.04$   & $2\times10^{-8}\pm4\times10^{-9}$     & 0.01/-\\
               & $\bar t$                      &$0.129\pm0.006$  & $0.260\pm0.012$  & $1387\pm70$  & $1.01\pm0.004$  & $1.85\times10^{-8}\pm5\times10^{-9}$  & 0.09/-\\
               & ${^3_\Lambda \bar H}$         &$0.128\pm0.005$  & $0.240\pm0.010$  & $1400\pm82$  & $1.0001\pm0.0005$& $5\times10^{-5}\pm4\times10^{-6}$    & 28/25\\
               & ${^3_{\bar\Lambda} \bar H}$   &$0.128\pm0.006$  & $0.240\pm0.012$  & $1400\pm100$ & $1.001\pm0.0004$& $4.5\times10^{-5}\pm5\times10^{-6}$   & 22/25\\
               & ${^3 He}$                     &$0.129\pm0.004$  & $0.220\pm0.011$  & $1400\pm100$ & $1.001\pm0.005$ & $2.7\times10^{-5}\pm6\times10^{-6}$   & 35/25\\
               & $\bar{^3 He}$                 &$0.128\pm0.006$  & $0.218\pm0.012$  & $1387\pm70$  & $1.0065\pm0.0004$ & $2.6\times10^{-5}\pm5\times10^{-6}$ & 21/25\\
\hline
\end{tabular}%
\end{center}}
\end{table*}

In Fig. 3, panel (a) shows the correlation of $T_0$ and $\beta_T$, while in panel (b) is the correlation
of $T_0$ and $V$. One can see that there is a negative correlation for each particle in both panel (a)
and (b). Panel (a) suggests that thermal motion ($T_0$) is transformed into collective motion ($\beta_T$)
as the system cools down. While panel (b) suggests that there is a strong squeeze during the collision
and naturally the parameter $V$ decreases, and it corresponds to higher degree of excitation of the system
which results in larger $T_0$.

To see the average transverse momentum $<p_T>$ of the particles and initial temperature $T_i$ of the system and
check their dependence on $m_0$ and collision energy, we presented
Fig. 4. $<p_T>$ and $T_i$
are calculated from the fit function over a given $p_T$ range of 0 to 4.5 GeV/\textit{c}, and then analyzed
the root mean square $\sqrt{<p^2_T>}$ over $\sqrt{2}$ ($\sqrt{<p^2_T>/2}$) which is the initial temperature of
the interacting system according to the
string percolation model \cite{56, 57, 58}.
It should be noted that although $T_i$ is directly related to $\sqrt{<p^2_T>/2}$ which we can obtain from
the $p_T$ spectra, but we obtain $T_i$ as usual in order to see its trend. Our discussion of $T_i$ is very
useful because it gives the understanding of the degree of excitation of the system in the initial state.
At the same time, we can also compare $T_i$ and $T_0$ to see the difference of temperature in the
system evolution.

The dependence of $<p_T>$ on $m_0$ and collision energy is displayed in panel (a). The symbols from left to right displays the
dependence of $<p_T>$ on the $m_0$, while from up to downwards is
its dependence on energy. Different symbols are used for the representation of different
energies. The symbols with upper filled portion represents
the particles, while down filled symbols represents the anti-particles.
It can be seen that $<p_T>$ increases with the particle mass. The more massive the particle is, the larger is the $<p_T>$.
In addition, due to isospin symmetry the $<p_T>$ of $t$, $^3_\Lambda H$, and $^3{He}$ and
their anti-particles are the same. $<p_T>$ is also observed to increase with increasing the
collision energy. The higher the collision energy is, the larger the energy transferred to the system is, which results in larger $<p_T>$.

Panel (b) represents the dependence of $T_i$ on $m_0$ and collision energy. We find that $T_i$ is mass dependent. The larger the mass of the particle is, the higher is the $T_i$. It is also observed that $T_i$ increase with increasing the collision energy. In addition, the $T_i$ from the spectra of $t$, $^3_\Lambda H$, and $^3{He}$ and their anti-particles is the same due to isospin symmetry. There is no coalescence effect on the $T_i$ is observed and the reason behind this is the different phenomenon in the initial and the final state of collisions.

\begin{figure*}[htb!]
\begin{center}
\hskip-0.153cm
\includegraphics[width=15cm]{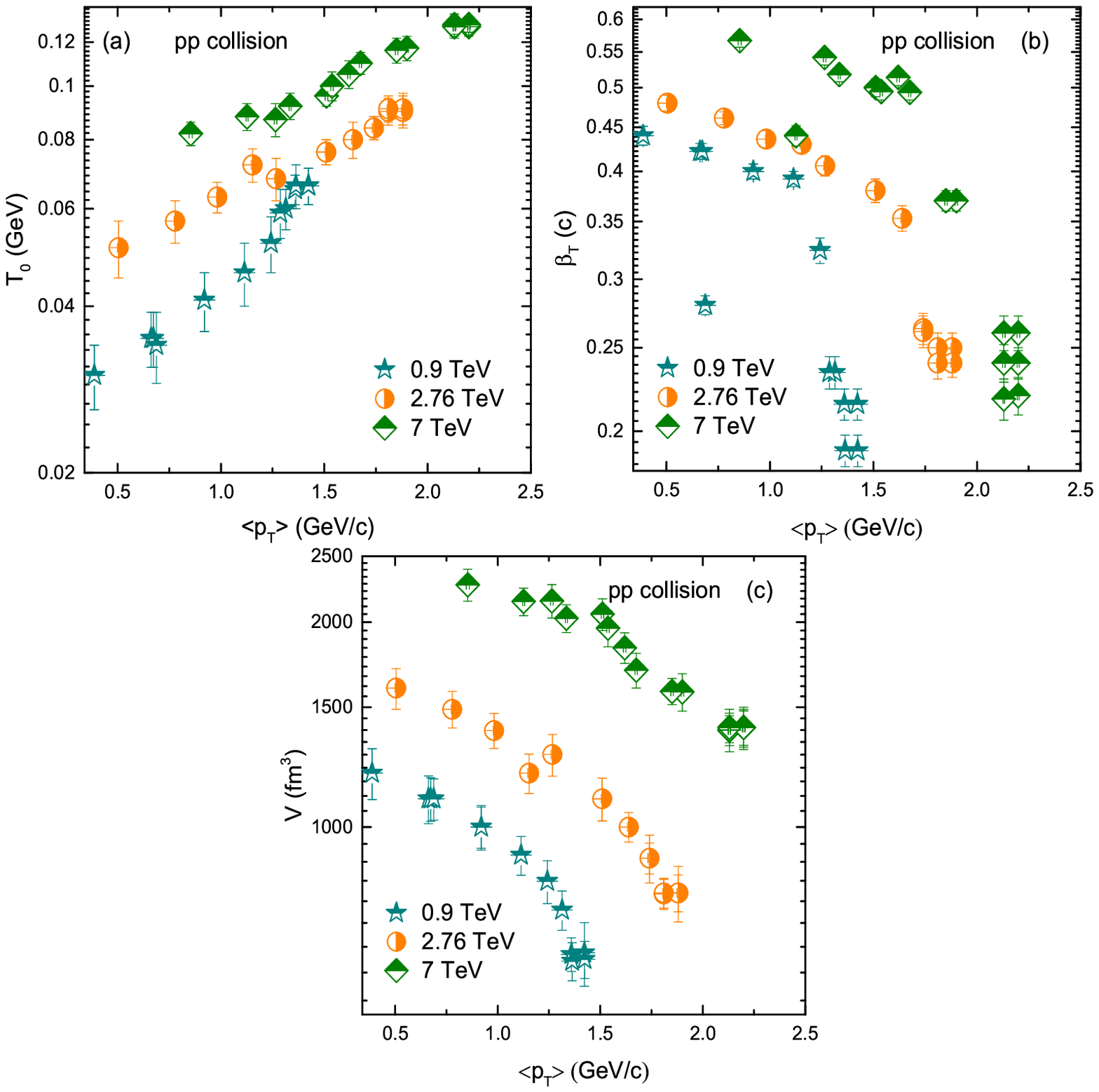}
\end{center}
Fig. 5. Correlation of (a) $T_0$ and $<p_T>$, (b) $\beta_T$ and $<p_T>$, and (c) $V$ and $<p_T>$ .
\end{figure*}

Figure 5 displays the correlation between $T_0$ and $<p_T>$, $\beta_T$ and $<p_T>$, and $V$ and $<p_T>$.
Panel (a) shows the positive correlation of $T_0$ and $<p_T>$, while panel (b) and (c) shows the negative correlation
among $\beta_T$ and $V$ with $<p_T>$. $T_0$ and $V$ are the parameters which are directly related to the freeze-out of the particles. Thus the positive correlation of $T_0$ and $<p_T>$ suggests that larger $<p_T>$ (energy) transfer to the system due to intense squeeze corresponds to higher degree of excitation of the system, and hence larger $T_0$. While on the other hand, the correlation is negative among $V$ and $<p_T>$. This suggests that the large
$<p_T>$ means large energy deposition in the system, which is possible, because the squeeze is intense, the system is dense and it will naturally leads to smaller $V$.

Fig. 5(b) illustrates the fact that there is a negative correlation between parameters $\beta_T$ and $<p_T>$. As discussed above, that there is a negative correlation between the parameters $T_0$ and $\beta_T$ of function in Eq. (1), which is due to conservation of particle energy, consisting of thermal energy part and kinetic energy of collective motion. Indeed, as seen from Figs. 5(a) and 5(b), on the whole $T_0$ increases while $\beta_T$ decreases with increasing $<p_T>$ of particles. As observed from Fig. 5(b), there is a clear separation among $\beta_T$ values of particles produced in proton-proton collisions at $\sqrt{s}$ = 0.9, 2.76, and 7 TeV. The largest and lowest $\beta_T$ values are observed in proton-proton collisions at the largest $\sqrt{s}$ =7 TeV and lowest $\sqrt{s}$ = 0.9 TeV collision energies, respectively, while $\beta_T$ values for intermediate energy $\sqrt{s}$ =2.76 TeV come in between. This can be explained by that the larger collision energies result in higher pressure gradients in collision zone and, hence, larger velocities of produced particles flying out from this zone at expansion stage.

Before going to conclusions we would like to clarify that our results are in agreement with the picture of hydrodynamic evolution. Compared with particles, light nuclei are emitted without being excited in the evolution process due to their massive mass. Obviously, light nuclei are produced earlier than particles in the process and correspond to higher temperature and smaller volume of the system. Because of the massive mass and higher temperature, light nuclei have larger mean transverse momentum. In addition, because of larger inertia, light nuclei have smaller transverse flow velocity.

\section{Summary and Conclusions}

We summarize here our main observations and conclusions.

(a) We studied the transverse momentum spectra of different particles and light nuclei
in $pp$ collisions at different center-of-mass energies and extracted the bulk properties,
including kinetic freeze-out temperature, transverse flow velocity and kinetic freeze-out volume.

(b) We observed that kinetic freeze-out temperature, transverse flow velocity and
kinetic freeze-out volume are mass dependent. The former increases as the particle mass
increases, while the latter two decrease with the increase of particle mass. The dependence
of kinetic freeze-out temperature and volume renders multiple kinetic freeze-out scenario.

(c)  We also observed isospin symmetry on the kinetic
freeze-out temperature and kinetic freeze-out volume, where triton, hyper-triton and helion,
and anti-triton, anti-hyper-triton and anti-helion freeze-out at the same time.

(d) We extracted the mean transverse momentum and initial temperature by an indirect method
and observed that they increase with mass, and have isospin symmetry.

(e) All the extracted parameters (kinetic freeze-out temperature, kinetic freeze-out volume, transverse flow velocity, initial temperature and mean transverse momentum) are observed to increase with increasing the center of mass energy due to the system getting higher degree of excitation.
\\
\\
{\bf Acknowledgments}
This work is supported by the
National Natural Science Foundation of China (Grant
Nos. 11875052, 11575190, and 11135011). We would
also like to acknowledge the support of Ajman
University Internal Research Grant NO. [DGSR Ref.
2021-IRG-HBS-12].
\\
\\
{\bf Author Contributions} All authors listed have made a
substantial, direct, and intellectual contribution to the work and
approved it for publication.
\\
\\
{\bf Data Availability Statement} This manuscript has no
associated data or the data will not be deposited. [Authors'
comment: The data used to support the findings of this study are
included within the article and are cited at relevant places
within the text as references.]
\\
\\
{\bf Compliance with Ethical Standards}
\\
\\
{\bf Ethical Approval} The authors declare that they are in
compliance with ethical standards regarding the content of this
paper.
\\
\\
{\bf Disclosure} The funding agencies have no role in the design
of the study; in the collection, analysis, or interpretation of
the data; in the writing of the manuscript, or in the decision to
publish the results.
\\
\\
{\bf Conflict of Interest} The authors declare that there are no
conflicts of interest regarding the publication of this paper.
\\
\\

\end{document}